\documentclass[final,3p,times,twocolumn]{elsarticle}

\usepackage{amssymb, amsmath}

\usepackage{hyperref}

\usepackage{color}
\usepackage{array}
\usepackage[final]{listings}

\definecolor{dark-gray}{gray}{0.15}
\definecolor{light-gray}{gray}{0.8}
\definecolor{lighter-gray}{gray}{0.9}

\definecolor{dark-green}{rgb}{0,0.4,0}
\definecolor{dark-red}{rgb}{0.2,0,0}

\newcolumntype{x}[1]{>{\centering\let\newline\\\arraybackslash\hspace{0pt}}m{#1}}

\journal{SoftwareX}

\newcommand{\rmd}{\mathrm{d}}
\newcommand{\nind}[1]{{(#1)}}
\newcommand{\Python}{Python}

\newcommand{\figref}[1]{Fig.~\ref{fig:#1}}

\begin{document}

\begin{frontmatter}



\title{PySpike -- A Python library for analyzing spike train synchrony}


\author{Mario Mulansky\corref{cor1}}
\ead{mario.mulansky@isc.cnr.it}

\author{Thomas Kreuz\corref{cor2}}
\ead{thomas.kreuz@cnr.it}

\cortext[cor1]{Corresponding author}

\address{Institute for Complex Systems, CNR, Via Madonna del Piano 10 -- 50019 Sesto Fiorentino, Italy}

\begin{abstract}
Understanding how the brain functions is one of the biggest challenges of our time.
The analysis of experimentally recorded neural firing patterns (spike trains) plays a crucial role in addressing this problem.
Here, the PySpike library is introduced, a Python package for spike train analysis providing parameter-free and time-scale independent measures of spike train synchrony.
It allows to compute similarity and dissimilarity profiles, averaged values and distance matrices.
Although mainly focusing on neuroscience, PySpike can also be applied in other contexts like climate research or social sciences.
The package is available as Open Source on Github and PyPI.

\end{abstract}


\end{frontmatter}


\section{Introduction}
\label{sec:introduction}
Gaining insight into the inner workings of the brain remains a largely unsolved challenge that requires combined efforts of biophysics, medicine, experimental as well as computational neuroscience~\cite{QuianQuiroga13}.
The basis for scientific advancement in this field are experimental recordings of neural activity usually represented in terms of spike trains, i.e.\ lists of spike times for each recorded neuron.
With sophisticated modern recording techniques, it is now possible to perform highly parallel measurements of neural activity, typically resulting in very large sets of spike trains~\cite{doi:10.1021/nn4012847,10.3389/fncom.2013.00137}.
This generates an increased demand for powerful and high quality data analysis tools that are capable of processing large data sets as produced by parallel recordings.


There exist numerous methods to analyze spike train data, e.g. based on spike count distributions, interspike intervals or exact spike times.
One very important approach is to quantify the synchrony between spike trains.
In the past decades several synchrony measures have been proposed~\cite{Victor96,VanRossum01,QuianQuiroga02b} which have already been used, among others, to quantify the reliability of neuronal responses~\cite{Tiesinga08}, to analyze the role of spike synchronization in feature binding~\cite{Singer09}, and to distinguish different stimuli in the context of neuronal coding~\cite{QuianQuiroga13}.

\begin{figure}[t]
 \centering
 \includegraphics[width=0.8\linewidth]{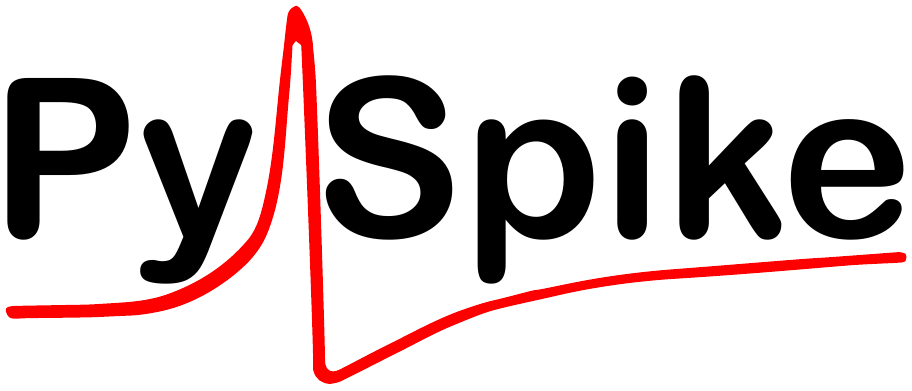}
 \caption{Logo of the PySpike library. \label{fig:logo}}
\end{figure}

The PySpike library\footnote{\url{www.pyspike.de}} introduced here (logo shown in \figref{logo}) is a Python package that allows one to compute two different dissimilarity measures, the ISI-distance~\cite{Kreuz07a}, the SPIKE-distance~\cite{Kreuz13} and additionally the similarity measure SPIKE-Synchronization~\cite{Kreuz15,Mulansky15}.
Each of these three methods is \emph{time-resolved}, \emph{parameter-free} and \emph{timescale independent} and therefore highly versatile.
Being time-resolved, for example, means these measure can detect changes in synchrony over time, while being parameter-free makes them readily applicable with unambiguous results, as no parameter optimization is required.
These measures have already been applied in many experimental studies in the past, for example~\cite{Dodla09,Ibarz10,Wildie12,DiPoppa13}.

Although developed with a neuroscientific context in mind, the synchrony measures discussed here can be applied to any form of discrete time series consisting of event sequences of any kind.
In fact, such measures have already been utilized in several other research areas, such as climate research~\cite{Malik10} or social sciences~\cite{Williams12,Rabinowitch15}.

\begin{figure*}[t]
 \centering
 \includegraphics[width=0.7\linewidth]{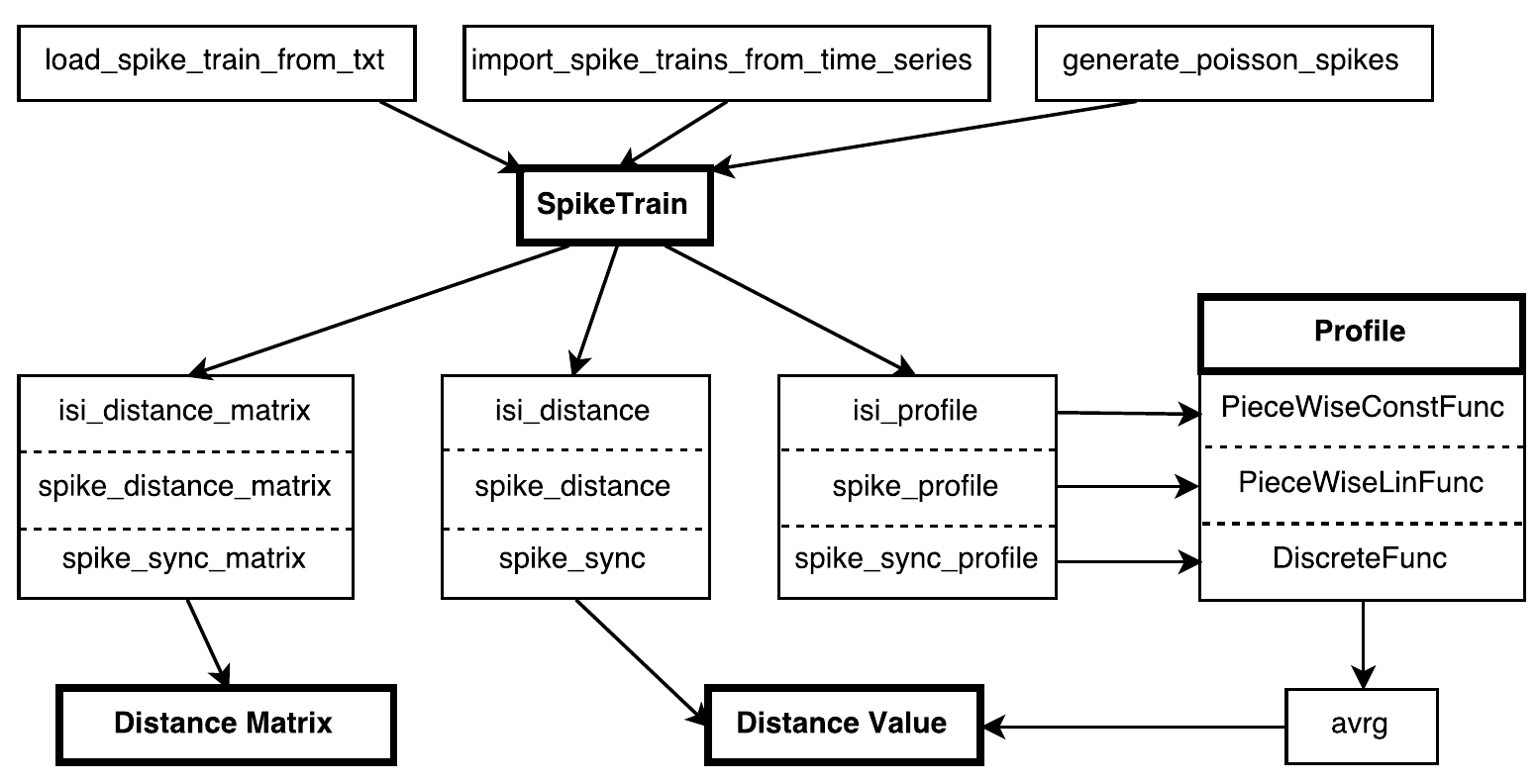}
 \caption{Structure of the PySpike package. The \textbf{SpikeTrain} is the central class for which several functions are provided to compute the dissimilarity or similarity-profiles, as well as distance/similarity values and matrices. \label{fig:structure}}
\end{figure*}

PySpike is a library aimed to perform automatized data analysis with Python scripts.
It is therefore a complementary approach to the SPIKY software package\footnote{\url{http://www.fi.isc.cnr.it/users/thomas.kreuz/Source-Code/SPIKY.html}}, a Matlab framework for spike train analysis providing a similar functionality but additionally offering a sophisticated GUI~\cite{Kreuz15,Bozanic14}.
Several other software packages for spike train analysis have been developed in the recent past, notably SyncPy\footnote{\url{https://github.com/syncpy/SyncPy}}~\cite{varni2015syncpy}, a Python based GUI for quantifying synchrony in time series.
However, it currently does not include the synchrony measures implemented in PySpike.
A C++ implementation of the spike train distance measures\footnote{\url{https://github.com/modulus-metric}} was presented in~\cite{Rusu14}, but it is based on sampled data and therefore of substantially inferior performance~\cite{Kreuz15}.
Finally, a comprehensive collection of scientific software for spike train analysis is also provided as part of~\cite{grun2010analysis}, aiming specifically at multivariate recordings.\footnote{\url{http://spiketrain-analysis.org/software}}

\section{Spike train distances}
Here, discrete time series are represented by \emph{spike trains}, sequences of time points denoting the occurrence of an event (spike) at those time points: $s = \{t_1, t_2, t_3, ...\}$.
Generally, a time-resolved distance measure maps a pair of spike trains $s_1$, $s_2$ onto a profile ${\{s_1, s_2\}\rightarrow S(t)}$ with ${0\leq S(t)\leq 1}$.
The overall distance value can easily be obtained by integration: ${D_S = \int S(t)\,\rmd t}$.

PySpike provides three such distance measures: ISI-distance, SPIKE-distance and SPIKE-Synchronization.
These methods are sensitive to different aspects of spike train synchrony (interspike intervals, exact spike timings, spike matching, respectively).
Hence, the choice of method should be informed by assumptions on how information is encoded in the spike trains.
In the following, we give a brief introduction to the measures provided in the PySpike library.
For a detailed discussion of the methods and their properties see \ref{app:math} and~\cite{Mulansky15}.

The \textbf{ISI-distance} profile $I(t)$, introduced in~\cite{Kreuz07a}, quantifies dissimilarity in terms of the relative differences of the concurrent interspike intervals of the two spike trains.
Essentially, it measures the relative differences of the instantaneous rates of the two spike trains, but it is not sensitive to exact spike timings.
The ISI-distance profile is a bivariate piecewise constant function.

The \textbf{SPIKE-distance} profile $S(t)$, first introduced in~\cite{Kreuz11} and refined in~\cite{Kreuz13}, represents a dissimilarity profile based on exact spike timings.
Thus, the SPIKE-distance quantifies spike train dissimilarity in terms of deviations from exact coincidences of spikes in the two spike trains.
This results in a bivariate piecewise linear profile.

While the fundamental definition of both the ISI- and the SPIKE-distance profile is bivariate (distance profile of two spike trains), the generalization to a multivariate context is a straightforward average over all spike train pairs~\cite{Kreuz09}.
%


\textbf{SPIKE-Synchronization}~\cite{Kreuz15, Mulansky15}
is a straight-forward, normalized coincidence counter with an adaptive coincidence window.
It quantifies similarity in terms of the fraction of coincidences between two spike trains and hence is a very intuitive measure.
The generalization of SPIKE-Synchronization for many spike trains can be defined based on all spike train pairs leading to a consistent multivariate framework with similarity again quantified as the overall fraction of coincidences in all spike trains~\cite{Kreuz15, Mulansky15}.

\section{Package Structure} \label{sec:structure}

\subsection{The Spike Train}
The central data structure of the PySpike library is the \lstinline+SpikeTrain+, a \Python\ class representing an individual spike train.
This class contains the (sorted) spike times as a \lstinline+numpy.array+ as well as the start and end time of the spike train.
Such \lstinline+SpikeTrain+ objects can either be created directly by providing the spike times, generated randomly from a Poisson process using the \lstinline+generate_poisson_spikes+ function, imported from text files via \lstinline+load_spike_train_from_txt+ or imported from time series via \lstinline+import_spike_trains_from_time_series+.
These objects then serve as input to calculate the distance measures, cf.~
\figref{structure}.

\begin{lstlisting}[
 float,
 basicstyle=\ttfamily\footnotesize,
 caption={Using profile and distance functions.},
 label={lst:isi_profile}]
# generate three arbitrary spike trains
# with start/end time of 0 and 4
st1 = SpikeTrain([1.0,2.0,3.0], edges=[0,4])
st2 = SpikeTrain([0.5,3.0,3.5], edges=[0,4])
st3 = SpikeTrain([2.5,3.8], edges=[0,4])

# bivariate profile of spike trains 1,2:
isi_prof = isi_profile(st1, st2)

# distance value as average over profile
isi_dist = isi_prof.avrg()

# faster: directly from the spike trains
isi_dist = isi_distance(st1, st2)

# profile of 3 spike trains:
spike_prof = spike_profile([st1, st2, st3])

# computing pairwise spike sync matrix
# of the whole list of spike trains:
spike_trains = [st1, st2, st3]

M = spike_sync_matrix(spike_trains)
# M is a 3x3 numpy array
\end{lstlisting}

\subsection{Computing Profiles}
Being time-resolved is a main advantage of the three spike train synchrony measures discussed here.
Hence, PySpike contains functionality to compute synchrony profiles: \lstinline+isi_profile+ computes the piecewise constant ISI-profile $I(t)$, \lstinline+spike_profile+ returns the piecewise linear SPIKE-profile $S(t)$ and \lstinline+spike_sync_profile+ yields the discrete SPIKE-Synchronization profile $C_k$.
As the three profiles have different mathematical properties, they are represented by different \Python\ classes in PySpike: \lstinline+PieceWiseConstFunc+, \lstinline+PieceWiseLinFunc+ and \lstinline+DiscreteFunc+, as seen in \figref{structure}.
By introducing specific data structures for each of these objects, PySpike can utilize their mathematical properties and use a highly efficient implementation while still providing a convenient user interface.
All three function classes provide an \lstinline+avrg+ member function that returns the total (time-averaged) distance of the respective measure.
In Listing~\ref{lst:isi_profile}, several profiles are computed and averaged.
Furthermore, the profiles offer a \lstinline+get_plottable_data+ member function for easy visualization of the profile, shown examparily in Listing~\ref{lst:plotting}.
%

\subsection{Computing Distances and Distance Matrices}

\begin{figure}[t]
 \centering
 \includegraphics[width=\linewidth]{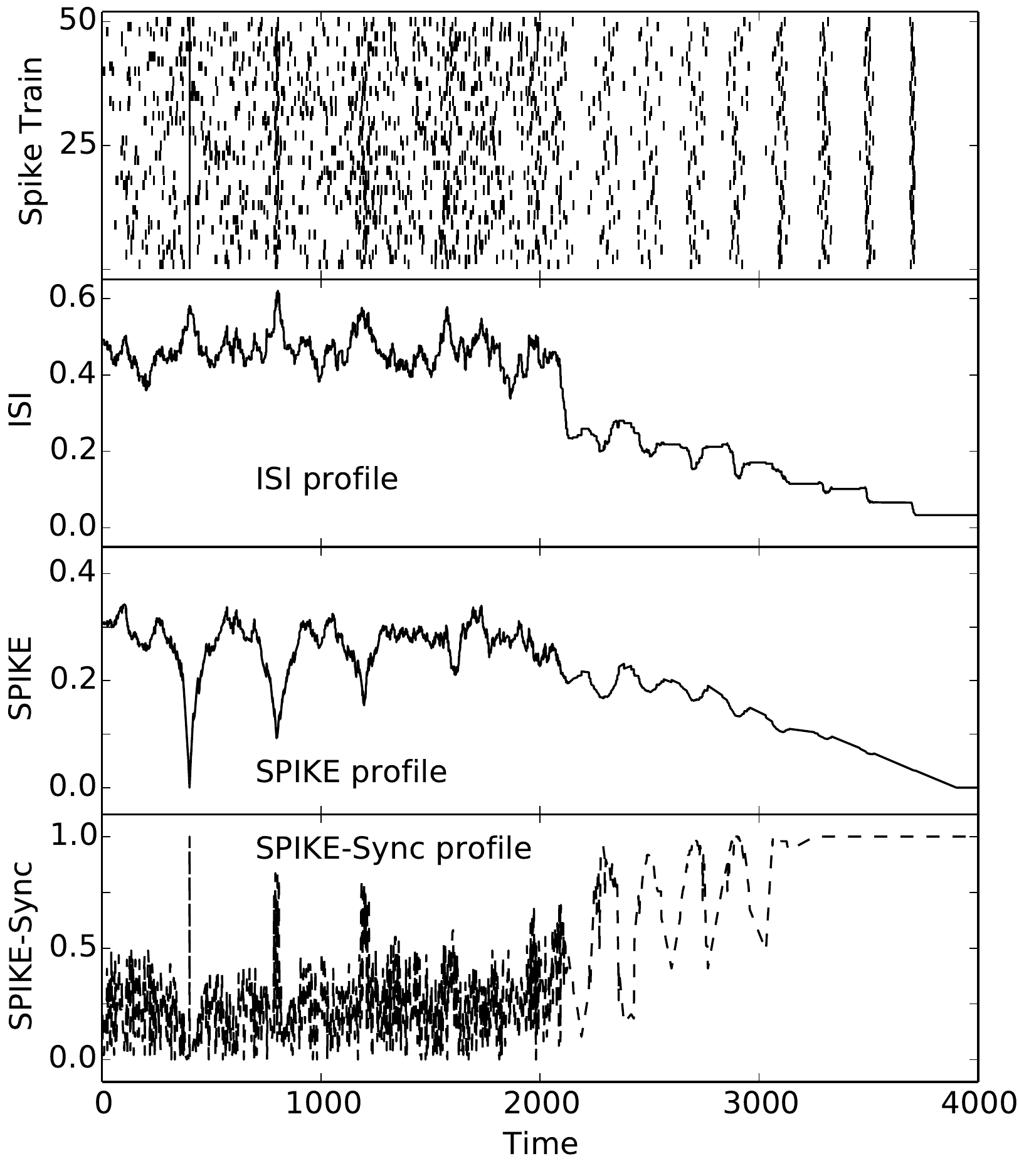}
 \caption{Multivariate ISI, SPIKE and SPIKE-Synchronization profiles for $M=50$ spike trains (shown in top panel). In this example, artificially generated spike trains are used where the first half consists of noisy spiking superimposed with a few synchronous events with more and more jitter. The second half contains increasingly synchronized events without the noisy background.
 Adapted from~\cite{Mulansky15}. \label{fig:profiles}}
\end{figure}

Besides calculating the time-resolved profiles, PySpike also provides functionality to directly compute the overall distance values for the ISI-distance and the SPIKE-distance as well as the overall SPIKE-Synchronization value.
The respective functions are \lstinline+isi_distance+, \lstinline+spike_distance+ and \lstinline+spike_sync+,
see Listing~\ref{lst:isi_profile} for an example on how to compute profiles and \figref{profiles} for example profile plots.
Although the overall distances can also be computed from the profiles by using the \lstinline+avrg()+ member function, the specific distance functions allow for a more efficient implementation and hence a significantly better performance, as will be discussed in Section~\ref{sec:performance}.

When analyzing sets of spike trains, one might not only be interested in the multivariate distance of the whole set, but also in the pairwise distances between all spike trains.
For this purpose, PySpike offers the \lstinline+isi_distance_matrix+, \lstinline+spike_distance_matrix+ and the \lstinline+spike_sync_matrix+ functions.
Listing~\ref{lst:isi_profile} shows an example on how to compute the SPIKE-Synchronization matrix. 
As all distance measures in PySpike are symmetric, the distance matrices are also always symmetric.

\begin{figure}[t]
 \centering
 \includegraphics[width=\linewidth]{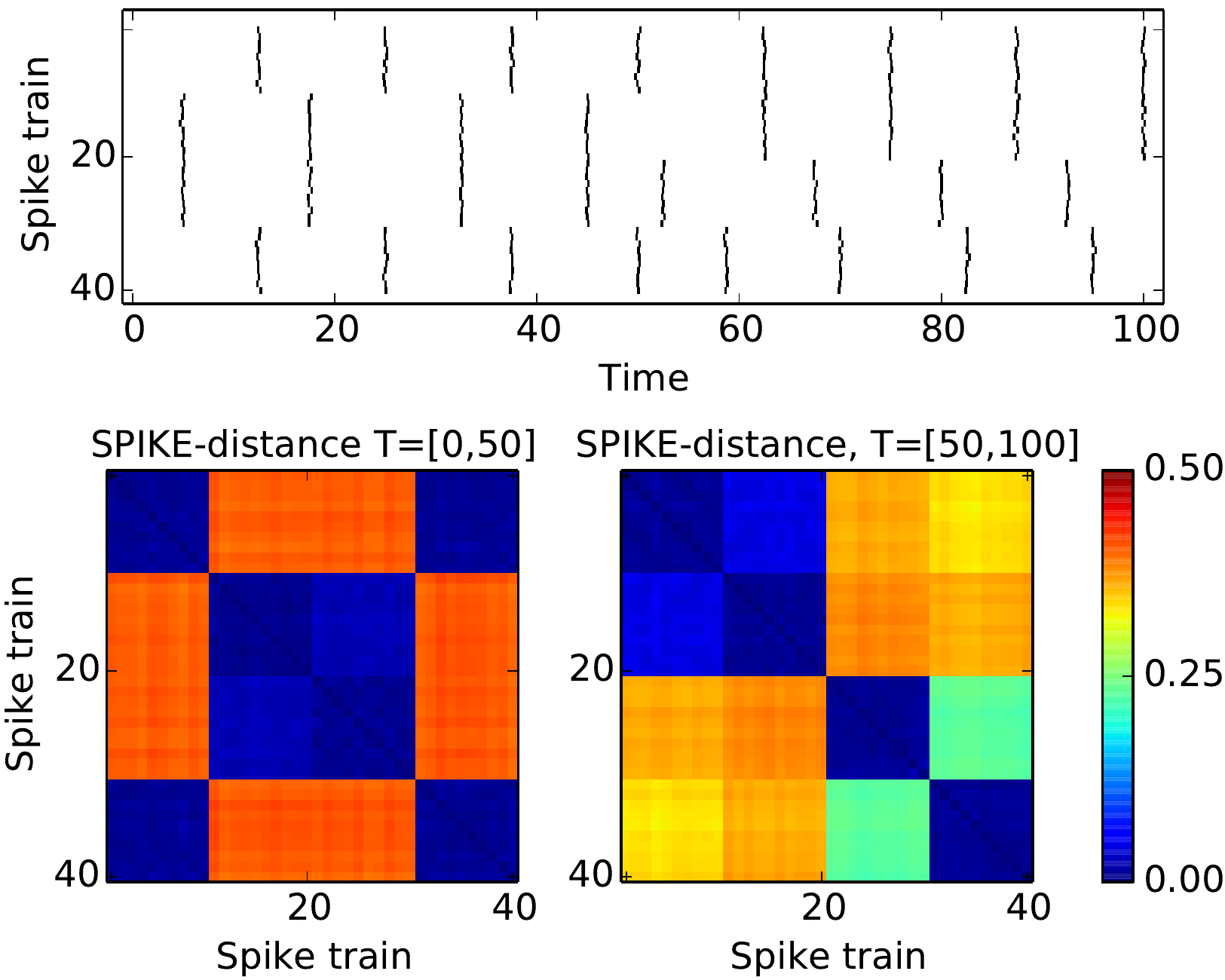}
 \caption{Spike-distance matrices for a set of $M=40$ artificial spike trains (top panel) with changing synchronous firing. The spike trains are divided into four groups, which fire in two different synchrony patterns in the first half and second half of the observation interval. This change of synchrony structure is clearly captured by using selective averaging for the SPIKE-distance matrices ($T=[0,50]$ on the left and $T=[50,100]$ on the right).\label{fig:matrix}}
\end{figure}

Finally, note that the distance and distance matrix functions can also be used to compute selective averages, that is averages over specified intervals of the profiles.
This is accomplished by providing the (optional) named parameter \lstinline+interval+ to the distance functions, and similarly for the \lstinline+avrg+ member function of the profiles.
An example for computing and plotting such a selective distance matrix is given in Listing~\ref{lst:plotting}.
This allows for a very detailed analysis of the profiles, for example by comparing the average distances before and after repeated stimuli or triggers.
\figref{matrix} shows an example of a changing synchrony pattern revealed by distance matrices with different averaging intervals.

\begin{lstlisting}[
 float,
 basicstyle=\ttfamily\footnotesize,
 caption={Load spike trains and plot profile and distance matrix.},
 label={lst:plotting}]
# load spike train data
spike_trains = load_spike_trains_from_txt(
                   "spike_trains.txt",
                   edges=(0, 4000))

# compute and plot ISI profile
isi_prof = isi_profile(spike_trains)
x, y = isi_prof.get_plottable_data()
plot(x, y, '-k')

# plot SPIKE distance matrix
# with selective averaging t=0...1000
spike_mat = spike_distance_matrix(
                spike_trains,
                interval=[0, 1000])
imshow(spike_mat)
\end{lstlisting}

%

%

\section{Implementation and Performance}

\subsection{Code Base}

PySpike is written following modern coding standards and best practices in scientific computing~\cite{10.1371/journal.pbio.1001745}, specifically adhering to the the official Style Guide for Python Code (PEP8)\footnote{\url{https://www.python.org/dev/peps/pep-0008/}}.
It is compatible with both the Python2 and Python3 runtime and available as open source software distributed under the BSD License.
The frontend is entirely implemented in Python, whereas for the crucial computations a Cython~\cite{behnel2010cython} version is provided for maximal performance, as shown in Section~\ref{sec:performance} below.
As most scientific Python libraries, PySpike requires NumPy as \lstinline+numpy+ arrays~\cite{Walt11} are the underlying data structures.
Furthermore, it is designed to interact with \lstinline+matplotlib+ for the plotting of profiles and distance matrices and it can also be easily combined with other scientific libraries such as \lstinline+scikit-learn+ for cluster analysis~\cite{scikit-learn}.

The PySpike library is developed on Github\footnote{\url{https://github.com/mariomulansky/PySpike/}} with periodic releases on PyPI\footnote{\url{https://pypi.python.org/pypi/pyspike/}}.
It is fully unit tested using the \lstinline+nosetest+ framework\footnote{\url{https://nose.readthedocs.org/en/latest}}, including an extensive test matrix on a continuous integration server~\footnote{\url{https://travis-ci.org/mariomulansky/PySpike}}, where all tests are performed on each commit using Python versions 2.6, 2.7, 3.3, 3.4, 3.5, and for both the Python and the Cython backend.

\subsection{Performance} \label{sec:performance}

With the increasing experimental abilities in neuroscience it is now possible to obtain parallel recordings of thousands and more spike trains in both cultured neurons~\cite{B907394A} and \emph{in-vivo} recordings~\cite{doi:10.1021/nn4012847}.
This 
requires an efficient implementation of spike train analysis tools that allow one to process such huge data sets within acceptable times.
However, Python is known for its very poor performance compared to low-level languages such as C/C++.
We therefore designed the PySpike library to consist of two parts: (1) A front-end implemented fully in Python representing the interface of the PySpike functionality to the library users; and (2) A backend providing the numeric implementation of the distance measures.
For the backend, PySpike provides two versions, a pure Python implementation and a much faster C implementation based on Cython.
The Cython backend is the default choice, and only if Cython is not available PySpike falls back to the Python backend.

\begin{figure}[t]
 \centering
 \includegraphics[width=\linewidth]{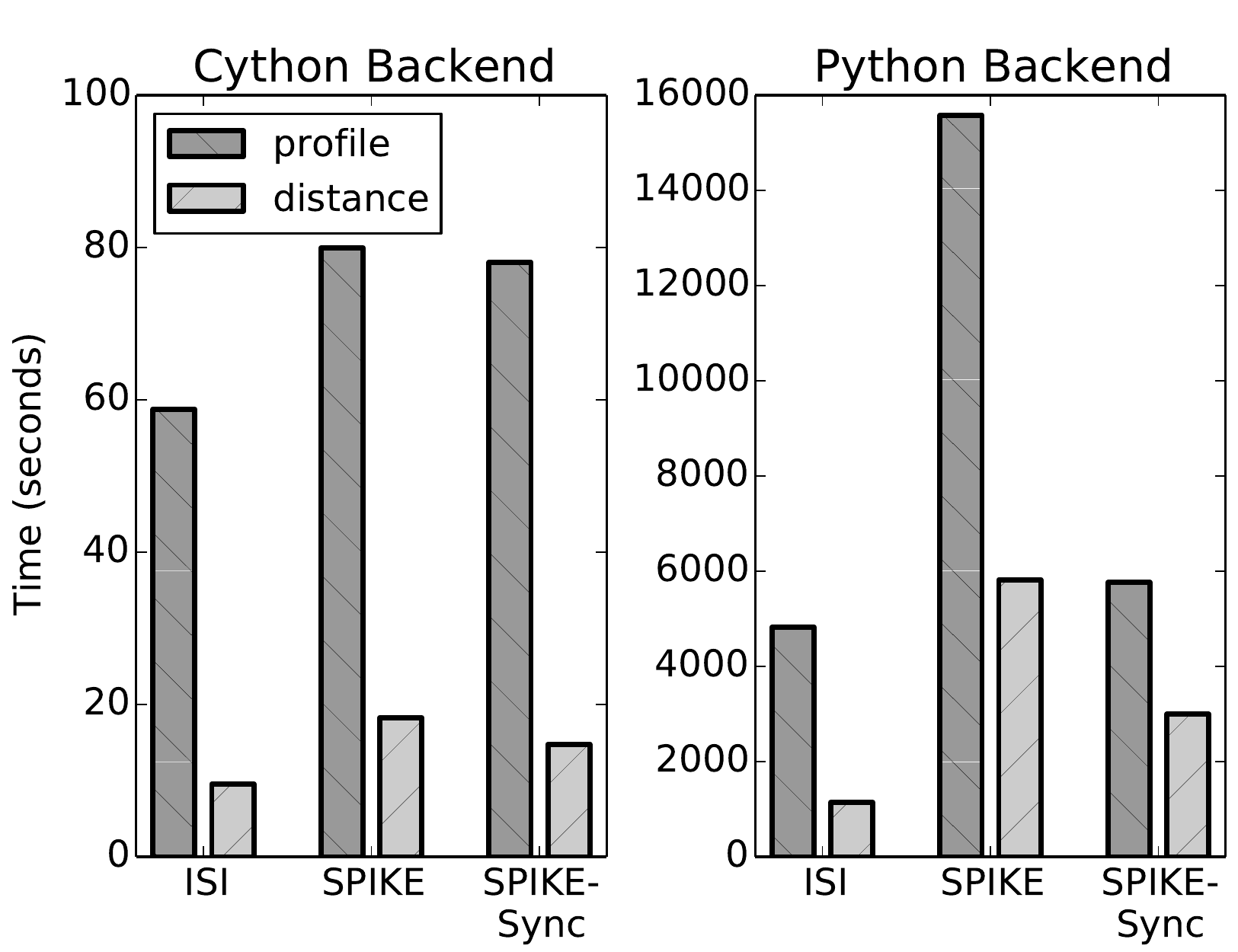}
 \caption{Runtime for computing the multivariate ISI-distance, SPIKE-distance and SPIKE-Synchronization for a set of $M=1000$ spike trains each containing $N=500$ spikes on average. Dark gray bars represent the runtime for first computing profiles and then averaging, while light gray bars indicate the direct computation of the distance values. Left panel shows the runtime (in seconds) for the Cython backend, while the right panel presents the runtime for the Python backend.\label{fig:perf}}
\end{figure}

Utilizing Cython (and therefore the speed of C), we were able to gain performance improvements of more than a factor of 200.
\figref{perf} shows the runtime of a multivariate analysis of $M=1000$ Poissonian spike trains each containing $N\approx500$ spikes, representative of a realistic experimental situation.
For both backends, Cython (left) and Python (right), we show the runtime of a full multivariate computation of the ISI-distance, SPIKE-distance and SPIKE-Synchronization for first computing the profile and then averaging, as well as for the specific distance functions.
All performance measurements where done on an Intel Core i5-3210M CPU @ 2.50GHz.
First, note that with the Python backend, the computation time reaches several hours, which is clearly beyond any tolerable runtime for a data analysis procedure.
Using the Cython backend, however, the computations require only 10-20 seconds (distance) or around one minute (profiles), which we believe are acceptable runtimes for a data analysis of this size.
Second, as mentioned earlier, the direct computations of the overall distance values (light gray in \figref{perf}) is significantly faster than first computing the profile and then averaging (dark gray), which is the reason why PySpike provides this functionality.

While \figref{perf} only shows the results for one size of dataset, namely $M=1000$ spike trains with each on average containing $N\approx500$ spikes, it is known that the runtime scales linearly with the number of spikes $N$ and quadratically with the number of spike trains $M$ (as the number of pairs grows $\sim M^2$).
Hence, the expected runtime for larger (or smaller) datasets can easily be estimated from the results in \figref{perf}.

\section{Conclusions and Outlook}

We presented PySpike, a Python library for measuring synchrony in experimental and simulated spike train data.
PySpike provides three parameter-free, time-scale independent and multivariate synchrony measures as well as facilities for plotting, Poisson spike train generation and selective averaging.
It is implemented in a clean, fully documented code base that follows official Python code style guidelines.
Furthermore, PySpike's development utilizes advanced software engineering techniques such as version control, unit testing and continuous integration.
By virtue of a Cython backend, it also shows excellent performance and hence is suitable for the analysis of large data sets.

Besides regular maintenance and bug fixing, future work on PySpike will include the addition of new synchrony measures specifically designed for spike trains with bursts~\cite{Raisanen16_prep}; support for event detection (transformation from continuous to discrete data); and extended functionality of spike train generation.
Current work is focused on the development of a spike order indicator that allows to analyze propagation patterns by quantifying temporal leader/follower properties within sets of spike trains~\cite{spike_order_wip}.
A preliminary Python implementation already exists and will become part of an official PySpike release in the near future.

In conclusion, we are convinced that the PySpike library is a helpful software package for spike train analysis.
It provides a clean, well documented and consistent Python interface to compute measures of spike train synchrony, while offering excellent performance.

\section*{Acknowledgements}
We thank N.~Bozanic and E.~R\"ais\"anen for numerous useful discussions.
Furthermore, we thank I.~Gnatenko, I.~Samuel and R.~Tomsett for their contributions to PySpike.
M.~M.\ and T.~K.\ acknowledge funding support from the European Commission through the Marie Curie Initial Training Network ``Neural Engineering Transformative Technologies (NETT)'' Project 289146, and T.~K.\ was also supported through the Marie Curie European Joint Doctorate ``Complex Oscillatory Systems: Modeling and Analysis (COSMOS)'' Project 642563.
Furthermore, T.~K.\ acknowledges the Italian Ministry of Foreign Affairs regarding their support of the Joint Italian-Israeli Laboratory on Neuroscience.

\appendix

\section{Mathematical Definitions} \label{app:math}

\small

\subsection{ISI-Distance} \label{app:isi}
\begin{figure}
\centering
 \includegraphics[width=0.49\textwidth]{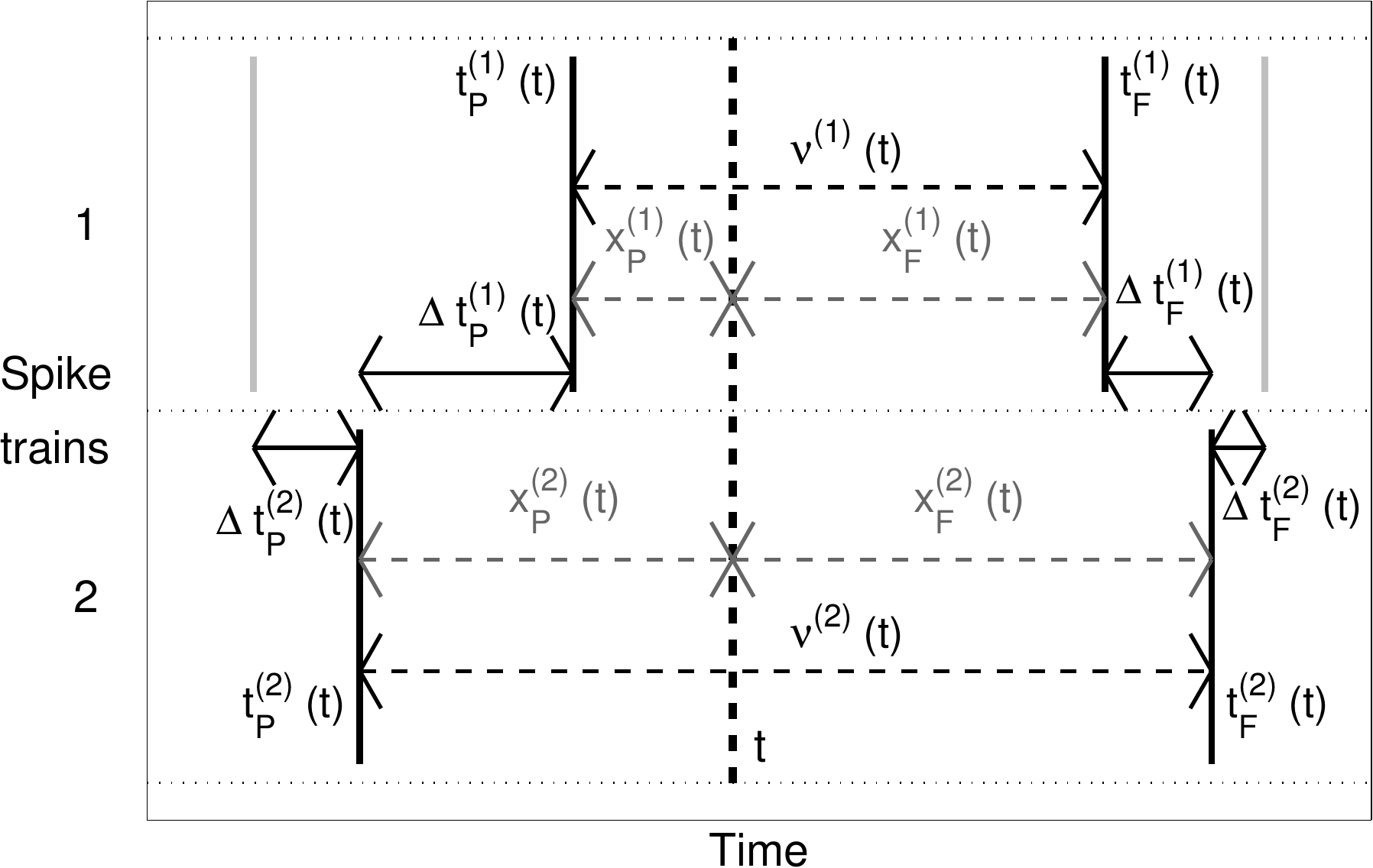}
\caption{Local definitions of interspike intervals and time differences required for the calculation of the ISI- and the SPIKE-profile.
\label{fig:spike_definitions}}
\end{figure}

With $\{t^\nind{1}_i\}$ being the spike times of the first spike train, its interspike intervals are given as ${\nu^\nind{1}_i = t^\nind{1}_{i+1}-t^\nind{1}_i}$ and similarly for the second spike train (cf.\ \figref{spike_definitions}).
The ISI-profile is then computed as the normalized absolute difference of these interspike intervals:
\begin{equation}
 I(t) = \frac{|\nu^{(1)}(t) - \nu^{(2)}(t)|}{\max\{\nu^{(1)}(t),\nu^{(2)}(t)\}},\quad t \in [T_0,T_1],
\end{equation} 
where $T_0 \leq t^\nind{1,2}_i \leq T_1$ are the edges of the spike trains.

\subsection{SPIKE-Distance} \label{app:spike}
The computation of $S(t)$ is based on the four corner spikes surrounding the current time $t$: the preceding spikes $t_P^{(1),(2)}(t)$ and the following spikes $t_F^{(1),(2)}(t)$ of each spike train (cf.\ \figref{spike_definitions}).
For each of the corner spikes, the distance to the closest spike of the \emph{other spike train} is computed, e.g.:
\begin{equation}
 \Delta t_P^\nind{1}(t) = \min_i\{|t_P^\nind{1}-t^\nind{2}_i|\},
\end{equation} 
and similarly for $\Delta t_P^\nind{2}$ and $\Delta t_F^{\nind{1},\nind{2}}$.
These distances are then combined by a linear interpolation according to the current time point $t$ resulting in the following quantity:
\begin{equation} \label{eqn:spike_weighted_dist}
 S_1(t) = \frac{\Delta t_P^\nind{1}(t) x_F^\nind{1}(t) + \Delta t_F^\nind{1}(t) x_P^\nind{1}}{\nu^\nind{1}(t)},
\end{equation} 
and similarly $S_2(t)$ is defined for the second spike train.
Based on these quantities, we finally arrive at the definition of the SPIKE-profile using the following normalization:
\begin{equation} \label{eqn:spike_profile}
 S(t) = \frac{S_1(t) \nu^\nind{2}(t) + S_2(t) \nu^\nind{1}(t)}{\frac12(\nu^\nind{1}(t)+\nu^\nind{2}(t))^2}, \quad t \in [0,T].
\end{equation}
Note, that this is a piecewise linear function as $S_{1,2}$ are piecewise linear while the other terms are piecewise constant.

\subsection{SPIKE-Synchronization} \label{app:spike-sync}
For SPIKE-Synchronization, a coincidence indicator~$C^{\nind{1},\nind{2}}_i$ is defined for every spike of the two spike trains $s^{\nind{1},\nind{2}}$, where $C_i=1$ if the spike at $t_i$ is part of a coincidence, and $C_i=0$ if not.
A coincidence is defined in terms of an \emph{adaptive} coincidence window $\tau$ according to the local firing rate:
\begin{equation} \label{eqn:spike_sync_tau}
 \tau^{\nind{1,2}}_{ij} = \frac12\min\{\nu^\nind{1}_{i}, \nu^\nind{1}_{i-1}, \nu^\nind{2}_{j}, \nu^\nind{2}_{j-1}\},
\end{equation} 
with $\nu^{\nind{1},\nind{2}}$ being the interspike intervals as above.
The value of the coincidence indicator is then given by:
\begin{equation} \label{eqn:spike_sync_cn}
 C^\nind{1}_i = \begin{cases} 1\quad\text{if}\quad \min_j(|t^\nind{1}_i - t^\nind{2}_j|) < \tau^{\nind{1,2}}_{ij}\\
        0\quad\text{otherwise}.
       \end{cases}
\end{equation} 
In the same way, the coincidence indicator for the second spike train $C^\nind{2}_i$ is computed.

The SPIKE-Synchronization profile is then given by the discrete function defined from the pooled coincidence indicators ${\{C_k\} = \{C^\nind{1}_i\}\cup\{C^\nind{2}_i\}}$ and spike times: ${\{t'_k\} = \{t^\nind{1}_i\}\cup\{t^\nind{2}_i\}}$.
Since integrals are not defined for discrete functions, the overall SPIKE-Synchronization value~$\text{SYNC}$ is obtained via summation:
\begin{equation} \label{eqn:overall_spike_sync}
 \text{SYNC} = \frac1M \sum_{k=1}^M C_k = \frac C M,
\end{equation}
where $M$ is the total number of spikes in the pooled spike train~$\{t'_k\}$ and $C$ denotes the total number of coincident spikes.
In contrast to the other two measures, SPIKE-Synchronization quantifies similarity rather than distance.

\normalsize
\section*{References}
\bibliographystyle{elsarticle-num} 
\bibliography{Kreuz_Bibliography,pyspike_bibliography}

\end{document}